\documentstyle[12pt]{article}

\input{epsf.tex}

\newcommand{\E}{{\cal{E}}}
\newcommand{\s}{\sigma}
\renewcommand{\d}{{\rm d}} 

\newcommand{\be}{\begin{equation}}      
\newcommand{\ee}{\end{equation}}
\newcommand{\bea}{\begin{eqnarray}}      
\newcommand{\eea}{\end{eqnarray}}
\def\J#1#2#3#4{(#1). {#2} {\bf #3}, #4}
\def\PTP{\em Prog. Theor. Phys.}
\def\PRL{\em Phys. Rev. Lett.}
\def\PRD{{\em Phys. Rev.} D}
\def\PR{\em Phys. Rev.}
\def\GRG{\em Gen. Relat. Grav.}
\def\APL{\em Ann. Physik}
\def\AJ{\em Ap. J.}
\def\JMP{\em J. Math. Phys.}
\def\CQG{\em Class. Quantum Grav.}
\def\ZP{\em Z. Phys.}
\def\PLA{\em Phys. Lett. A}

\begin{document}
\title{Remarks on the Charged, Magnetized\\ 
Tomimatsu-Sato $\delta=2$ Solution}
\author{O.~V.~Manko\dag ,
V.~S.~Manko\ddag ~and J.~D.~Sanabria-G\'omez\ddag}
\date{}
\maketitle

\vspace{-1cm}

\begin{center}
\dag Physics Faculty, Lomonosov Moscow State University,\\
Moscow 119899, Russian Federation\\
\medskip
\ddag Departamento de F\'\i sica,\\ Centro de Investigaci\'on y de
Estudios Avanzados del IPN,\\ 
A.P. 14-740, 07000 M\'exico D.F., Mexico
\end{center}

\vspace{.1cm}

\begin{abstract}
The full metric describing a charged, magnetized generalization of the
Tomimatsu-Sato (TS) $\delta=2$ solution is presented in a concise explicit
form. We use it to investigate some physical properties of the solution;
in particular, we point out the existence of naked ring singularities in
the hyperextreme TS metrics, the fact previously overlooked by 
the researchers, and we also demonstrate 
that the ring singularities can be eliminated by
sufficiently strong magnetic fields in the subextreme case, while in the
hyperextreme case the magnetic field can move singularities to the 
equatorial plane.
\end{abstract}


\newpage

\noindent{\bf 1. INTRODUCTION}

\vspace{.5cm}

In \cite{MMS} we have presented an exact asymptotically flat 4-parameter
solution of the Einstein-Maxwell equations constructed with the aid of
Sibgatullin's method \cite{S} which generalizes the well-known
Tomimatsu-Sato $\delta=2$ metric \cite{TS1,TS2} and is defined by the
Ernst complex potentials $\E$ and $\Phi$ \cite{E} of the 
form\footnote{In what follows we write $\s$ instead of $\s^2$ used in
Ref.~1 to underline the fact that $\s$ may assume both positive and
negative values. Throughout the paper units are used in which the 
gravitational constant and the velocity of light are equal to unity} 
\bea
{\cal E}&=&\frac{A-2mB}{A+2mB}, \quad \Phi=\frac{2C}{A+2mB},\nonumber\\
A&=&(k^2x^2-\s y^2)^2-(k^2-\s)^2-2ik^3axy(x^2-1) \nonumber\\
&&-(1-y^2)[a(k^2-2\s)+2qc][a(y^2+1)+2ikxy], \nonumber\\
B&=&kx[k^2(x^2-1)+\s(1-y^2)]
-iy(1-y^2)[a(k^2-2\s)+2qc], \nonumber\\
C&=&k^2(x^2-1)(kqx+icy)+(1-y^2)\{ k(ac+q\s) \nonumber\\
&&-iy[aq(k^2-2\s)+c(2q^2-\s)]\}, \nonumber\\
\s&\equiv&c^2/(m^2-a^2-q^2), \quad k\equiv\sqrt{m^2-a^2-q^2+\s},
\eea
$x$, $y$ being the generalized spheroidal coordinates, and $m$, $a$, $q$,
$c$ being four arbitrary real parameters representing the mass, angular
momentum, charge and magnetic dipole moment.

The expressions for the corresponding metric functions $f$ and $\gamma$
entering the axisymmetric line element
\bea
\d s^2&=&k^2f^{-1}\Bigl[e^{2\gamma}(x^2-y^2)
\Bigl(\frac{\d x^2}{x^2-1}+\frac{\d y^2}{1-y^2}\Bigr) \nonumber \\
&&+(x^2-1)(1-y^2)\d\varphi\Bigr]-f(\d t-\omega\d\varphi)^2,
\eea
have been given in \cite{MMS} only in terms of the polynomials $A$, $B$
and $C$, while for the remaining function $\omega$ we have been forced
to give a rather cumbersome expression because of not having at hand a 
good strategy for the simplification of complicated coefficients 
involving four independent parameters.

Recently notwithstanding we have succeeded in getting very concise 
expressions for $f$ and $\gamma$, and a by far simpler expression for 
$\omega$ than in Ref.~1. The main 
objective of this paper, therefore, will be the
presentation of the simplest metric for the exterior field of a charged,
magnetized, spinning mass in a concise explicit form most suitable for 
its analysis and possible applications. Besides, we shall analyse some
physical properties of the metric, such as, e.g., the multipole moments,
singularities and limits. We shall demonstrate, in particular, that
the superextreme TS solutions do have naked ring singularities (contrary
to the statement made in \cite{Y1} on the absence of such singularities) 
which are located outside the equatorial plane, but can be moved to the
equatorial plane by an appropriate choice of the magnetic dipole
parameter. 

\vspace{.5cm}

\noindent{\bf 2. THE METRIC FUNCTIONS AND LIMITING CASES}

\vspace{.5cm}

The first step in our search for concise expressions of the metric 
coefficients $f$, $\gamma$ and $\omega$ corresponding to the solution (1) 
was based on Yamazaki's idea~\cite{Y1,Y2} to represent these coefficients 
in terms of the quantities $x^2-1$ and $1-y^2$. Posterior major
simplifications have come from the analysis of the factor structure of
the metric coefficients, and in that work the papers by 
Hoenselaers~\cite{H} and Perj\'es~\cite{P} have been our guides. A tedious 
but straightforward algebra with the use of the Mathematica computer 
programme~\cite{W} has finally led us to the following elegant 
expressions for $f$, $\gamma$ and $\omega$:
\bea
f&=&\frac ED, \quad e^{2\gamma}=\frac E{k^8(x^2-y^2)^4},
\quad \omega=-\frac{2(1-y^2)F}E, \nonumber \\
E&=&\{[k^2(x^2-1)+\s(1-y^2)]^2+a[a(d-\s)+2qc](1-y^2)^2\}^2
\nonumber\\
&&-4k^2(x^2-1)(1-y^2)[k^2a(x^2-y^2)+2(a\s-qc)y^2]^2,
\nonumber\\
D&=&\{(k^2x^2-\s y^2)^2+2kmx[k^2(x^2-1)+\s(1-y^2)]
\nonumber\\
&&+a[a(d-\s)+2qc](y^4-1)-d^2\}^2\nonumber\\
&&+4y^2\{ k^3ax(x^2-1)+[a(d-\s)+2qc](kx+m)(1-y^2)\}^2,
\nonumber\\
F&=&4k^2(x^2-1)[k^2a(x^2-y^2)+2(a\s-qc)y^2]
\nonumber \\
&&\times\{ kmx[k^2(x^2+1)-\s(y^2+1)]+k^2x^2(2m^2-q^2)-\s dy^2\}
\nonumber \\
&&-\{[k^2(x^2-1)+\s(1-y^2)]^2+a[a(d-\s)+2qc](1-y^2)^2\}
\nonumber \\
&&\times\{2k^2qc(x^2-y^2)+(1-y^2)[ad(2kmx+2m^2-q^2)
\nonumber \\
&&-(a\s-2qc)(2kmx+m^2+a^2)]\},
\nonumber \\
d&\equiv&m^2-a^2-q^2.
\eea

Note that the above expression for $F$ is five times shorter
than the respective expression in Ref.~1! Together with the formulae for
$E$ and $D$ it defines explicitly the simplest metric able to
describe the exterior field of a charged, magnetized, spinning mass.

Some properties of the solution (1)-(3) can be better seen if its
relativistic Simon's multipole moments \cite{S} are given. Below we
write out the first four moments obtainable from (1) with the aid of
the Hoenselaers-Perj\'es procedure \cite{HP}, $M_i$ describing the
distribution of the mass (Re$M_i$) and angular momentum (Im$M_i$), and
$Q_i$ describing the electric (Re$Q_i$) and magnetic (Im$Q_i$) fields:
\bea
&&M_0=2m, \quad M_1=4ima,\nonumber\\ 
&&M_2=-2m(m^2+3a^2-q^2-\s),\nonumber\\
&&M_3=-8ima(m^2+a^2-q^2-\s),\nonumber\\
&&Q_0=2q, \quad Q_1=2i(c+2aq),\nonumber\\
&&Q_2=-2[2ac+q(m^2+3a^2-q^2-\s)],\nonumber\\
&&Q_3=-2i[(m^2+a^2-q^2-\s)(c+aq)+2a^2c],
\eea
whence the asymptotic flatness of the solution follows immediately, as
well as the physical interpretation of the parameters $m$, $a$, $q$ and
$c$ as determining, respectively, the total mass, total angular momentum
per unit mass, total charge and magnetic dipole moment of the source.
The latter four physical quantities are not restricted anyhow, so that
the solution is equally applicable to both the sub- and superextreme
cases which correspond to the real and pure imaginary values of $k$,
respectively.

Let us consider now the limiting cases of the solution (1)-(3).

a) The limit $q=c=0$ leads to the Tomimatsu-Sato $\delta=2$ metric
\cite{TS1,TS2}.

b) Another well-known limit is Bonnor's two-parameter solution \cite{B}
for a static massive magnetic dipole ($q=a=0$).

c) The case $c=0$ corresponds to the charged version of the
Tomimatsu-Sato $\delta=2$ solution constructed by Ernst \cite{E2}.

d) When $q=0$, one arrives at the Manko-Ruiz solution \cite{MR} which
is a stationary generalization of the Bonnor metric \cite{B}.

e) When the parameters happen to satisfy the relation
\be
m^2-a^2-q^2-\s=0,
\ee
the solution can be interpreted as a specific electromagnetic
generalization of the Kerr metric \cite{K} different from the Kerr-Newman
spacetime \cite{N} if $\s\ne0$.

f) The limit $a=0$ is of special interest. It provides an exact 
three-parameter analog to the approximate solution used by Bonnor \cite{B2}
for his analysis of the dragging of inertial frames by a charged
massive magnetic dipole. In view of the potential physical importance
of this particular solution we write it out explicitly:
\bea
f&=&\frac ED, \quad e^{2\gamma}=\frac E{k^8(x^2-y^2)^4},
\quad \omega=\frac{4qc(1-y^2)F}E, \nonumber \\
E&=&[k^2(x^2-1)+\s(1-y^2)]^4-16k^2q^2c^2y^4(x^2-1)(1-y^2),
\nonumber\\
D&=&[(k^2(x^2-1)+\s(1-y^2)]^2[k^2(x^2+1)-\s(y^2+1)+2kmx]^2
\nonumber\\
&&+16q^2c^2y^2(kx+m)^2(1-y^2)^2,
\nonumber\\
F&=&[k^2(x^2-1)+\s(1-y^2)]^2[k^2(x^2-1)+(2kmx+k^2+m^2)(1-y^2)]
\nonumber \\
&&+4k^2y^2(x^2-1)(kmx+m^2-q^2)(k^2x^2-\s y^2+kmx),
\nonumber \\
k&\equiv&\sqrt{m^2-q^2+\s}, \quad \s\equiv\frac{c^2}{m^2-q^2}.
\eea

The total angular momentum of this solution is equal to zero, as well as
all its higher rotational multipole moments. At the same time, the metric
has a non-vanishing d$t$d$\varphi$ term characterized by the coefficient
$\omega$! According to Bonnor \cite{B2}, this effect of frame-dragging
by a charged, massive magnetic dipole is due to the Poynting vector which
produces flows of energy in the equatorial plane where the frame-dragging
occurs.

Note that the above metric cannot be obtained from the Bonnor magnetic
dipole solution \cite{B} by application of the Kramer-Neugebauer charging
transformation \cite{KN} since the latter transformation generates a
specific non-vanishing angular momentum which depends on the parameters
of charge and magnetic dipole moment \cite{KN,KSHM}.

As a final remark concluding this section let us point out that the
potentials (1) corresponding to the pure imaginary values of $k$ belong
to the Chen-Guo-Ernst family of electrovacuum hyperextreme solutions
\cite{CGE}.

\vspace{.5cm}

\noindent{\bf 3. SINGULARITIES AND STATIONARY LIMIT SURFACE}

\vspace{.5cm}

The structure of singularities of the electrovacuum rational function 
solutions has some similar as well as distinctive features compared with
the pure vacuum case, but in both cases the singularities arise as 
solutions of the equation
\be
A+2mB=0.
\ee

For the real-valued $k$ (the subextreme case) the solution (1)-(3),
similar to the TS $\delta=2$ solution, has two singular points on the
symmetry axis, $x=1$, $y=\pm1$, and besides may have a naked ring
singularity in the equatorial plane. However, if the latter ring
singularity is inevitably present in the subextreme TS $\delta=2$
solution, it is not necessarily the case for the charged, magnetized
TS $\delta=2$ solution (1)-(3) where the ring singularity can be
eliminated by the magnetic field. 

In the following three diagrams we have shown the location of 
singularities for particular values of the parameters of the subextreme 
TS $\delta=2$ solution (Figure~1.i) and of the solution (1)-(3) 
(Figures~1.ii,iii). In addition we have plotted there the shape of the
stationary limit surface for each case (on this surface defined by the
equation
\be
E=0
\ee
the time-like Killing vector becomes a null vector). Note that all 
figures have been plotted in the Weyl-Papapetrou cylindrical coordinates
$(\rho,z)$ introduced via the formulae
\be
x=\frac1{2k}(r_++r_-),\quad y=\frac1{2k}(r_+-r_-),\quad
r_\pm\equiv\sqrt{\rho^2+(z\pm k)^2}.
\ee

On can see (Figure~1.i) that all the three singularities of the
TS $\delta=2$ solution lie on the stationary limit surface (this is of
course the general property of the stationary vacuum solutions). In the
presence of the electromagnetic field with the value of the magnetic
dipole parameter less than the critical one, the naked ring singularity
is still present but it is not already located on the stationary limit
surface (Figure~1.ii). When the parameter $c$ exceeds the critical value 
(which is approximately 0.8254 for the chosen values of the 
parameters $m$, $a$ and $q$), the ring singularity disappears 
(Figure~1.iii).

\begin{figure}[htb]
\centerline{\epsfysize=45mm\epsffile{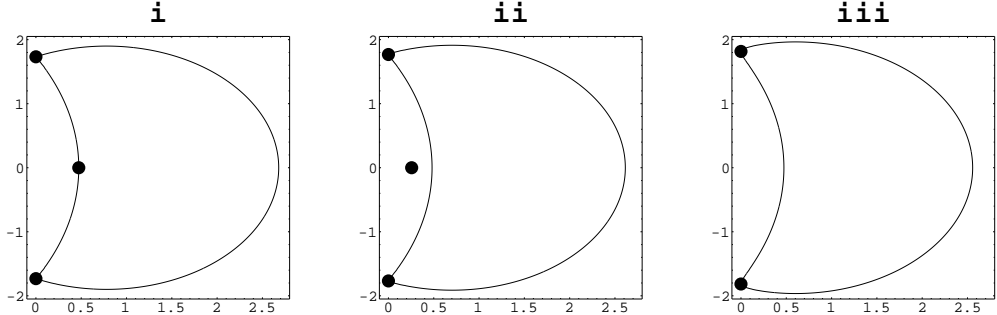}}
\caption{Ergosphere and singularities in the subextreme case. 
The particular choice of the parameters is:
i) m = 2, a = 1, q = c = 0; ii) m = 2, a = 1, q = 0.2, c = 0.7; 
iii) m = 2, a = 1, q = 0.2, c = 1.} 
\end{figure}

\begin{figure}[htb]
\centerline{\epsfysize=45mm\epsffile{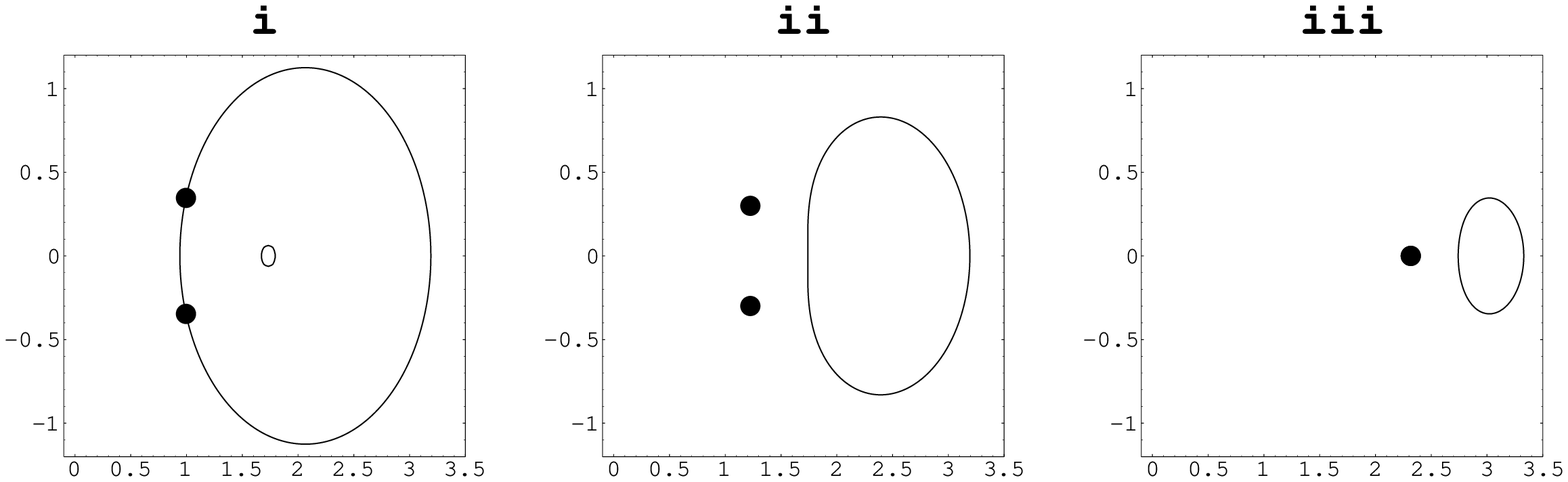}}
\caption{Ergosphere and singularities in the hyperextreme case. 
The particular choice of the parameters is:
i) m = 1, a = 2, q = c = 0; ii) m = 1, a = 2, q = 0.2, c = 1.1; 
iii) m = 1, a = 2, q = 0.2, c = 2.3.} 
\end{figure}

Turning now to the hyperextreme case characterized by the pure imaginary
values of $k$, it should be first of all remarked that Yamazaki's
statement (\cite{Y1}, p.~2505) about the absence of naked ring
singularities in the hyperextreme TS metrics is erroneous.  It is true
that the hyperextreme TS solutions have no singular points in the
equatorial plane; however, the singularities arise outside the equatorial
plane. Figure~2.i shows that the stationary limit surface of the
hyperextreme TS $\delta=2$ solution is a torus with two ring singularities
on it. By adding the electromagnetic field, the singularities can be
brought closer to each other (Figure~2.ii). Lastly, one can see that a
sufficiently strong magnetic field moves the singularities to the
equatorial plane (Figure~2.iii), eliminating one of them.

\vspace{.5cm}

\noindent{\bf 4. THE MAGNETIC POTENTIAL}

\vspace{.5cm}

The electric and magnetic fields in the solution (1)-(3) are described,
respectively, by the $A_4$ and $A_3$ components of the electromagnetic
four-potential. The component $A_4$ is simply the real part of the Ernst
potential $\Phi$ defined by (1). The determination of $A_3$ is most
simple via the construction of Kinnersley's complex scalar potential
$\Phi_2$ \cite{Kin}. The details of the derivation of the latter potential
in Sibgatullin's method can be found, e.g., in Ref.~23; hence, in what
follows we shall restrict ourselves to only writing out the resulting
expression for $\Phi_2$:
\bea
\Phi_2&=&\frac{2G}{A+2mB}-2iq, \nonumber \\
G&=&k^2(x^2-1)\{(1-y^2)[c(kx+3m)+iy(ac+q\s)]+2kaqx
\nonumber\\
&&+iqy[k^2(x^2+1)+2kmx-2\s]\}+(1-y^2)\{2(kx+m)
\nonumber\\
&&\times[aqd+c(2m^2+q^2)]-[aq(d-\s)+c(2q^2-\s)]
\nonumber\\
&&\times[(kx+m)(1-y^2)+2iay]+2m[c(\s-2a^2-q^2)-aq\s]\}.
\eea

The magnetic potential $A_3$ is determined consequently as the real part
of $\Phi_2$. On the four diagrams (Figures~3.i-iv) we have plotted the 
magnetic lines of force for different particular parameter sets which 
cover both the sub- and superextreme cases.

\begin{figure}[htb]
\centerline{\epsfysize=100mm\epsffile{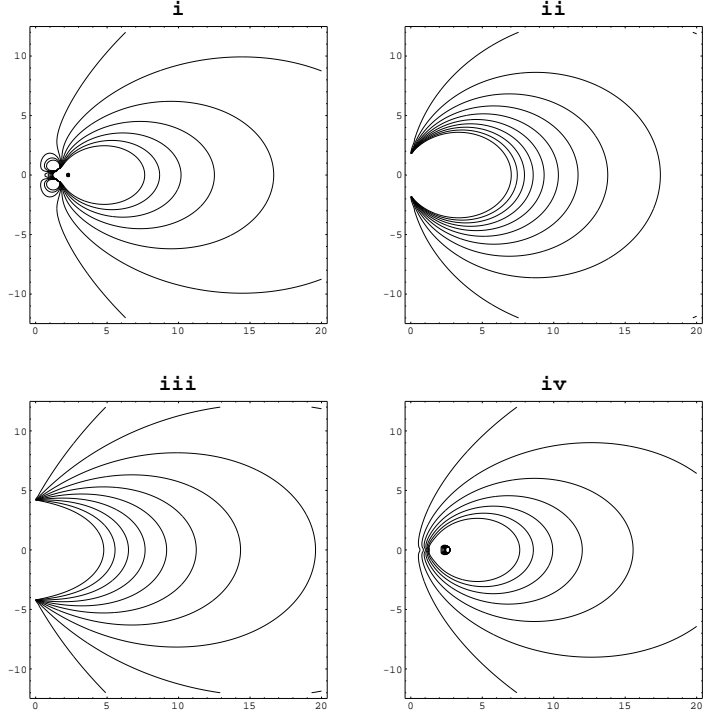}}
\caption{Magnetic lines of force. The particular choice of the
parameters is:
i) m = 1, a = 2.5, q = 0, c = 1; ii) m = 2, a = 1, q = -0.5, c = 1; 
iii) m = 1, a = 0, q = -0.2, c = 4; iv) m = 1, a = 1.5, q = -0.2,
c = 2.5.} 
\end{figure}

\vspace{.5cm}

\noindent{\bf 5. CONCLUSIONS}

\vspace{.5cm}

Therefore, we have succeeded in giving concise explicit expressions for
all the metric coefficients defining the charged, magnetized
generalization of the Tomimatsu-Sato $\delta=2$ solution, and we have
analysed some physical properties of the new electrovac solution. The
latter has been shown to have several well-known limits, as well as
some new limits among which the three-parameter solution for a charged,
massive magnetic dipole is probably of special interest.

The study of singularities of the solution (1)-(3) has enabled us to
correct an old erroneous belief that the hyperextreme TS solutions
have no ring singularities. At the same time we have shown that
a magnetic field can eliminate these naked ring singularities in the
subextreme case, and move them to the equatorial plane from outer
regions in the superextreme case.

The existence of naked ring singularities located outside
the equatorial plane in the hyperextreme TS solutions
probably makes the latters not quite appropriate for modelling the
exterior fields of single infinitesimally thin relativistic disks where 
the Neugebauer-Meinel (global) solution \cite{NM1,NM2} and the 
hyperextreme Kerr solution \cite{K} seem to be the only possibilities 
to describe such objects (however, the hyperextreme TS metrics most likely
could describe the fields of superposed disks). At the same time, there 
is of course no any problem in interpreting the TS metrics as 
describing the exterior fields of deformed masses, and besides we hope
that the subclasses of the magnetized hyperextreme TS solutions whose
singularities are located exclusively in the equatorial plane still 
could be considered as good candidates even for the description
of the exterior fields of single magnetized thin disks. 

\newpage

\noindent{\bf ACKNOWLEDGEMENTS}

\vspace{.5cm}

We would like to thank Prof.~J.~Pleba\'nski for stimulating and
interesting discussions. This work was supported by Project 26329-E 
from Conacyt, Mexico. J.D.S-G. also acknowledges financial 
support from Colciencias of Colombia and from SRE of Mexico.

\vfill\eject

\end{document}